\documentclass[letterpaper, 10 pt, conference]{ieeeconf}  % Comment this line out
\IEEEoverridecommandlockouts                              % This command is only
\usepackage{graphics} % for pdf, bitmapped graphics files
\usepackage{epsfig} % for postscript graphics files
\usepackage{mathptmx} % assumes new font selection scheme installed
\usepackage{times} % assumes new font selection scheme installed
\usepackage{amsmath} % assumes amsmath package installed
\usepackage{amssymb}  % assumes amsmath package installed
\usepackage{subfigure}
\usepackage{array}
\title{\LARGE \bf
A Packetized Direct Load Control Mechanism for Demand Side Management*
}

\author{Bowen Zhang$^{1}$ and John Baillieul$^{2}$% <-this % stops a space
\thanks{*The authors gratefully acknowledge support of the U.S. National Science Foundation under EFRI Grant 1038230}% <-this % stops a space
\thanks{$^{1}$Corresponding author. Division of Systems Eng., Boston University, 15 St. Mary’s St., Brookline, MA 02446, email:
        {\tt\small bowenz@bu.edu}}%
\thanks{$^{2}$Dept. of Electrical and Computer Eng., Dept. of Mechanical Eng., and Division of Systems Eng., Boston University, 110 Cummington St., Boston, MA 02215, email:
        {\tt\small johnb@bu.edu}}%
}

\begin{document}

\maketitle
\thispagestyle{empty}
\pagestyle{empty}

%%%%%%%%%%%%%%%%%%%%%%%%%%%%%%%%%%%%%%%%%%%%%%%%%%%%%%%%%%%%%%%%%%%%%%%%%%%%%%%%
\begin{abstract}

Electricity peaks can be harmful to grid stability and result in additional generation costs to balance supply with demand. By developing a network of smart appliances together with a quasi-decentralized control protocol, direct load control (DLC) provides an opportunity to reduce peak consumption by directly controlling the on/off switch of the networked appliances. This paper proposes a packetized DLC (PDLC) solution that is illustrated by an application to air conditioning temperature control. Here the term packetized refers to a fixed time energy usage authorization. The consumers in each room choose their preferred set point, and then an operator of the local appliance pool will determine the comfort band around the set point. We use a thermal dynamic model to investigate the duty cycle of thermostatic appliances. Three theorems are proposed in this paper. The first two theorems evaluate the performance of the PDLC in both transient and steady state operation. The first theorem proves that the average room temperature would converge to the average room set point with fixed number of packets applied in each discrete interval. The second theorem proves that the PDLC solution guarantees to control the temperature of all the rooms within their individual comfort bands. The third theorem proposes an allocation method to link the results in theorem 1 and assumptions in theorem 2 such that the overall PDLC solution works. The direct result of the theorems is that we can reduce the consumption oscillation that occurs when no control is applied. Simulation is provided to verify theoretical results.

\end{abstract}

\section{Introduction}
It is well known that the day-to-night electricity usage is oscillatory, with a usage valley appearing through the night and a peak occurring during the day. At the same time, high-frequency (minute-to-minute and faster) oscillation results from randomly occurring aggregations of individual loads with short duty cycle \cite{LKK}. The importance of reducing high-frequency peaks in usage is multi-fold. We can more easily maintain the stability of the grid with reduced amounts of generation reserves such that the grid frequency and voltage are stable. Generation cost can be reduced since we will not use generators with large marginal costs. Among all classes of electricity demand, thermostatic loads have been a major contributor to problems of high peak usage \cite{RV}. At the same time, thermostatic loads provide thermal capacity such that we can regulate their usage pattern as long as certain baseline thermal requirements are met. This paper presents an approach to carrying out such regulation by means of a novel information-based method for {\em direct load control}.

Historically, thermostatic loads (air conditioners, electric space heating systems, water heaters, etc.) have been operated in an uncoordinated
fashion resulting in the power grid being exposed to costly random load fluctuations. Taking note of the past decades's development of networked
control system technologies \cite{B1} and novel concepts enabled by smart appliances, such as the so-called {\em Internet of Things} \cite{INT}, we shall study the control of a local network of loads wherein the objective of control is to suppress spikes and fluctuations in usage. The approach uses real-time data from individual devices and local temperature sensors communicating with a central operator who distributes quantized amounts of energy to service the load demands according to a protocol for {\em direct load control} (DLC) that we shall describe below.

Various approaches have been proposed to formulate the DLC problem with the objective of peak load management. The load curve has been studied using a state-queueing model where thermal set points are adjusted automatically as a function of electricity price or outside temperature in \cite{LC} and \cite{LCW}. Dynamic programming has been applied in \cite{HC} to minimize the production cost in a unit commitment problem, and in \cite{LWCHC} to minimize the disutility of consumers resulting from DLC disruption. Monte Carlo simulation has been applied in \cite{RV} to evaluate the effectiveness of a specific DLC approach which minimized the discomfort of overall temperature deviation subject to constraints in transmission lines. Multi-server queueing theory has been used to calculate the mean waiting time of consumers when the usage authorization is limited during peak hours in \cite{LKK}. This system has been applied in a total of 449 residential units located in Seoul with good performance.

The objective of our approach is to monitor and control aggregate electricity use in order to avoid random spikes in demand that would otherwise occur. The mechanism that implements the approach is something that we call \textit{packetized} direct load control (PDLC). The term \textit{packetized} refers to the idea of \textit{time-packetized} energy where the central operator authorizes electricity usage of individual loads for a fixed amount of time $\Delta t$. After the elapse of time $\Delta t$, the central operator reschedules the authorization. For each building, the central operator is connected to the on/off switch of thermostatic loads (fan coils or room air conditioners). Users in the building are assumed to authorize the operator to control the on/off switch of their thermostatic smart appliances once they provide the operator their preferred temperature set point. The central operator, who receives thermal information on all the appliances at each decision instant, has the objective to maintain all appliances within their comfort band by selectively turning on or off these thermostatic loads at discrete time instants. The PDLC provides flexibility in adjusting building consumption since we are actually dealing with a discrete time decision making problem where the central operator schedules packets at the beginning of each interval. It will be shown that given a minimum critical level of energy capacity, it is possible to both eliminate demand peaks and guarantee a narrow comfort band around each consumer's preferred temperature setting. In a theoretical sense, it is further shown that the width of the comfort band can be made to approach zero by letting the packet length approach zero, although practically speaking the cycle time of an air conditioning unit cannot be made arbitrarily short. In the end, the PDLC solution is able to smooth the consumption oscillations, and this in turn enables buildings to consume smaller amounts of reserves dispatched from the ISO.

The remainder of the paper is organized as follows. Section \ref{two} introduces the set up of the PDLC mechanism, followed by the investigation of a thermal model in section \ref{three}. Section \ref{four} and \ref{five} discuss the transient and steady state operation of the PDLC solution respectively. Section \ref{link section} discusses an allocation solution to link theorem 1 and theorem 2. A robustness analysis is given in section \ref{six}. Section \ref{seven} provides simulation results. Section \ref{eight} concludes the paper and proposes future work.
\section{The PDLC Mechanism Setup}
\label{two}
This section describes the model in terms of which the PDLC mechanism is proposed. The following few points compose the background of the proposed approach.

(1) The PDLC controls the thermostatic loads in a building, such as air conditioners, refrigerators, and water heaters. The thermal dynamic model of these appliances does not differ much; the investigation of the thermal model of air conditioners in the next section can be extended to other thermostatic loads with minor change.

(2) The PDLC is assumed to be an on/off control. All the appliances are assumed to running with rated power if packet is authorized, or consume nothing if packet authorization is denied. There is no intermediate operation choice. 

(3) Different feeders are in charge of different types of loads, and they are all connected to the central operator who schedules electricity packets. The loads that have been grouped together in the same feeder consume energy at the same rates when they are operating. The overall consumption of the building is the sum of the consumption in each feeder controlled by the PDLC mechanism plus a certain portion of uncontrollable loads, including lighting and plug-in devices such as computers, televisions, and other small appliances. We assume that the consumption of the uncontrollable loads is independent of thermostatic loads and the environments (temperature, humidity, etc.), and these uncontrollable loads are subtracted from the analysis of the PDLC framework.

(4) It is assumed that the target level of consumption in each feeder of thermostatic load is available beforehand, which is defined as the average consumption during peak time when no control is applied. The value of the proposed method rests on the evidence-based assumption that the consumption curve without control would oscillate around the target level, and consumption peaks will frequently exceed the target level by a significant amount. The control objective of the PDLC is to make the consumption curve smoother around the target level with minimum oscillation.

\section{Air Conditioner Thermal Model}
\label{three}
\begin{figure}[htb]
\centering
\includegraphics[width=0.4\textwidth,height=0.2\textheight]{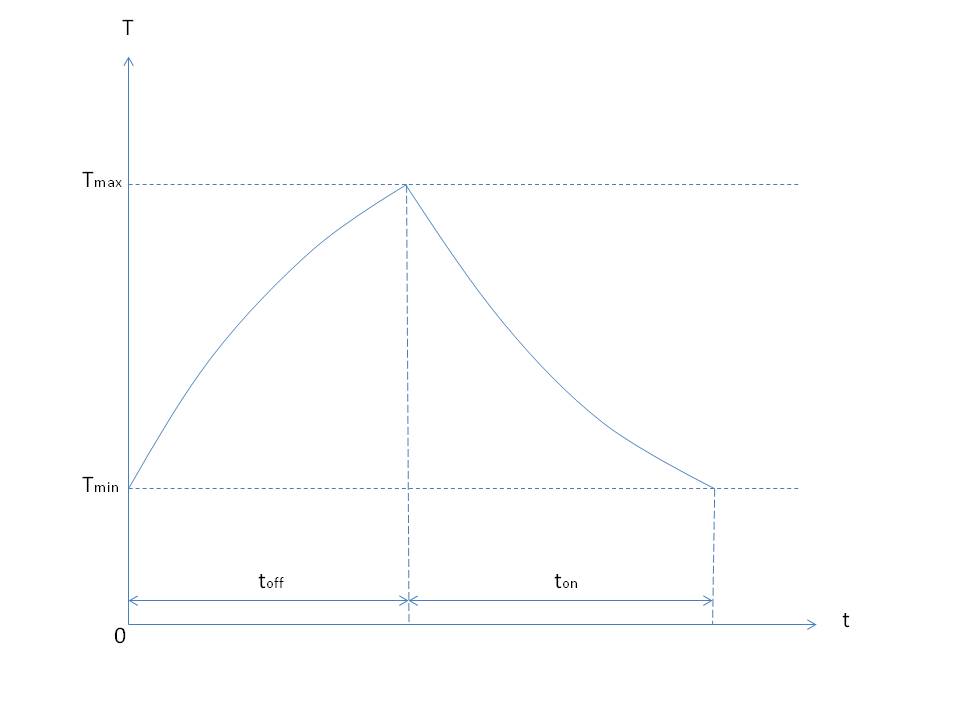}
\caption{Typical air conditioner duty cycle}
\label{duty}
\end{figure}

A model of the thermal dynamics of an air conditioner is developed as follows. Ihara and Schweppe presented a dynamic model for the temperature of a house regulated by air conditioning, and this has been shown to capture the behaviour of air conditioner loads accurately \cite{IS}. The temperature dynamics in continuous time (CT) is given by
\begin{equation}
\label{CT}
\frac{dT}{dt}=\frac{T_{out}-T-T_{g}u}{\tau},
\end{equation}
where $T_{out}$ is the outside temperate, $T_{g}$ is the temperature gain of air conditioner if it is on, $\tau$ is the effective thermal time constant of the room, and $u$ is binary valued specifying the state of thermostat. The unit of parameters is Fahrenheit as in the original paper. The temperature dynamic model in discrete time (DT) with interval $\Delta t$ is given by
\begin{equation}
\label{DT}
T_{k+1}=(1-a)T_{k}+aT_{out}-bu_{k},
\end{equation}
where $a=1-e^{-\frac{\Delta t}{\tau}}$, $b$=$aT_{g}$, and $u_{k}$ is $u$'s value during the $k$-th interval. We first derive the duty cycle off-time $t_{off}$ and on-time $t_{on}$ based on the CT model for the case in which there is no PDLC and the air conditioner is operating in the traditional way under the control of its own thermostat. $T_{max}$ and $T_{min}$ are the comfort band boundaries. To get $t_{off}$, we set $u=0$ in (\ref{CT}), which means that the air conditioner is turned off. Rearranging terms we have
\begin{equation}
\frac{dT}{dt}+\frac{1}{\tau}T-\frac{T_{out}}{\tau}=0,
\end{equation}
whose general solution is given by 
\begin{equation}
T(t)=Ce^{-\frac{t}{\tau}}+T_{out}.
\end{equation}
Since $t_{off}$ is the time that temperature arises from $T_{min}$ to $T_{max}$ in the case of traditional thermostat control, we choose initial condition $T(0)=T_{min}$ to solve for $t_{off}$. See Fig.\ref{duty}. $C=T_{min}-T_{out}$. The overall solution of temperature evolution is given then by
\begin{equation}
T(t)=(T_{min}-T_{out})e^{-\frac{t}{\tau}}+T_{out}.
\end{equation}
The value of $t_{off}$ would satisfy $T(t_{off})=T_{max}$. After calculation we will have
\begin{equation}
\label{TOFF}
t_{off}=\tau\ln\frac{T_{out}-T_{min}}{T_{out}-T_{max}}.
\end{equation}
Similarly we calculate $t_{on}$ when $u=1$,
\begin{equation}
\label{TON}
t_{on}=\tau\ln\frac{T_{max}+T_{g}-T_{out}}{T_{min}+T_{g}-T_{out}}.
\end{equation}

The traditional duty cycle dynamics characterized by $t_{off}$ and $t_{on}$ provide the baseline against which the PDLC protocol of the next section is evaluated. To evaluate the PDLC solution, we consider its transient and steady state operation. The next section will discuss its transient operation.

\section{Transient Operation of the PDLC}
\label{four}
The motivation of the PDLC solution is to allow buildings consume electricity at a level that minimizes oscillation close to a target. Denote the total number of consumers by $N_{c}$, the number of authorized packets by $m$, the set point in room $i$ by $T_{set}^{i}$, and the room temperature in room $i$ at time $k$ by $T_{k}^{i}$. The transient process is defined as the duration before the average room temperature converges to the average room set point $T_{set}^{ave}=\frac{1}{N_{c}}\sum_{i=1}^{N_{c}}T_{set}^{i}$. The theorem below provides a solution that guarantees the convergence of average room temperature under the assumption that $m$ packets are being allocated to a pool of appliances during each packet interval.

\textbf{Theorem 1.} If the fixed number of packets $m=N_{c}\frac{T_{out}-T_{set}^{ave}}{T_{g}}$ is used in each time interval $\Delta t$, then the average room temperature $T_{k}^{ave}=\frac{1}{N_{c}}\sum_{i=1}^{N_{c}}T_{k}^{i}$ converges to the average room set point $T_{set}^{ave}$.
 
\textit{Proof:} We use the DT model to derive the convergence of the average room temperature. According to (\ref{DT}), we can represent the number of authorized packets in terms of the DT model parameters $a$ and $b$ as follows
\begin{equation}
\label{MM}
m=N_{c}\frac{T_{out}-T_{set}^{ave}}{T_{g}}=N_{c}(T_{out}-T_{set}^{ave})\frac{a}{b}.
\end{equation}
In one packet length, the total temperature decrease $T_{k}^{dec}$ by $m$ packets is given by
\begin{equation}
\label{DEC}
T_{k}^{dec}=mb=N_{c}a(T_{out}-T_{set}^{ave}),
\end{equation}
where the last equality follows from (\ref{MM}). Similarly, the total temperature increase $T_{k}^{inc}$, which is caused by indoor/outdoor temperature difference, is given by
\begin{equation}
\label{INC}
T_{k}^{inc}=\sum_{i=1}^{N_{c}}a(T_{out}-T_{k}^{i})=N_{c}aT_{out}-a \sum_{i=1}^{N_{c}}T_{k}^{i}.
\end{equation}
The total temperature change $T_{k}^{ch}$ is given by
\begin{equation}
%\begin{array}{cll}
T_{k}^{ch} =  T_{k}^{inc}-T_{k}^{dec} =  aN_{c}(T_{set}^{ave}-T_{k}^{ave}).
%\end{array}
\end{equation}
$T_{k+1}^{ave}$ can be expressed recursively as
\begin{equation}
T_{k+1}^{ave}=T_{k}^{ave}+\frac{1}{N_{c}}T_{k}^{ch}=T_{k}^{ave}+a(T_{set}^{ave}-T_{k}^{ave}).
\end{equation}
We will have the difference between $T_{set}^{ave}$ and average room temperature at time $k+1$ given by
\begin{equation}
T_{set}^{ave}-T_{k+1}^{ave}=(1-a)(T_{set}^{ave}-T_{k}^{ave})=e^{-\frac{\Delta t}{\tau}}(T_{set}^{ave}-T_{k}^{ave}).
\end{equation}
For any small deviation $\epsilon>0$ from $T_{set}^{ave}$, we will have
\begin{equation}
|T_{set}^{ave}-T_{k}^{ave}|=e^{-\frac{k\Delta t}{\tau}}|T_{set}^{ave}-T_{0}^{ave}|<\epsilon,
\end{equation}
after $k$ steps, with $k$ satisfying
\begin{equation}
\label{STEP1}
k>\frac{\tau}{\Delta t}\ln\frac{|T_{set}^{ave}-T_{0}^{ave}|}{\epsilon}.
\end{equation}
This means the average room temperature will converge to an arbitrarily small neighbourhood of $T_{set}^{ave}$ after finite number of steps.$\blacksquare$

We say that the system is in \textit{Steady State Thermal Equilibrium} (SSTE) when the average room temperature is within a sufficiently small neighbourhood of $T_{set}^{ave}$. If the system is in SSTE at time $k^{\star}$, then the system will be in SSTE for $k\geq k^{\star}$ as long as we provide $m=N_{c}\frac{T_{out}-T_{set}^{ave}}{T_{g}}$ packets at each interval. According to (\ref{STEP1}), the convergence speed depends on $T_{0}^{ave}$ and $\tau$. If these two parameters do not provide a quick convergence with few steps ($T_{0}^{ave}$ being large in a warm load pick up process), we can adjust the number of packets as a function of the average temperature deviation $T_{k}^{ave}-T_{set}^{ave}$ at time $k$. Let the modified number of packets be given by
\begin{equation}
\label{revise number packet}
m=N_{c}\frac{T_{out}-T_{set}^{ave}}{T_{g}}[1+g(T_{k}^{ave}-T_{set}^{ave})],
\end{equation}
where $g$ is a non-negative coefficient. In this case, for any $\epsilon>0$  we can similarly prove that after $k^{'}$ steps the deviation of average room temperature from $T_{set}^{ave}$ is smaller than $\epsilon$, with $k^{'}$ satisfying
\begin{equation}
\label{STEP2}
k^{'}>-\ln[1-(1-e^{-\frac{\Delta t}{\tau}})G]\ln\frac{|T_{set}^{ave}-T_{0}^{ave}|}{\epsilon},
\end{equation}
where $G=1+g(T_{out}-T_{set}^{ave})$ can be understood as the convergence gain parameter. Comparing (\ref{STEP2}) with (\ref{STEP1}), we have $k^{'}<k$ for the same $\Delta t$ since $G>1$. The larger the value of $G$ (or $g$), the quicker the convergence. If $m$ in (\ref{revise number packet}) is not an integer, we can choose the ceil $\lceil m \rceil$ as the number of packets scheduled. The proof remains valid under this choice.

Theorem 1 indicates that the average consumption is proportional to the total population $N_{c}$ by the coefficient $\frac{T_{out}-T_{set}^{ave}}{T_{g}}$. The physical meaning of this coefficient is the thermostat mean status. Define 
\begin{equation}
\label{ONOFF}
s_{on}=\frac{T_{out}-T_{set}^{ave}}{T_{g}},s_{off}=1-s_{on},
\end{equation}
representing the mean on-status and off-status of the thermostat. These two variables will be used in the second theorem for the steady state analysis of the PDLC. Note that an essential implicit assumption is that $\frac{T_{out}-T_{set}^{ave}}{T_{g}}<1$, i.e. there is enough cooling capacity to serve the consumer population.

\section{Steady State Operation of the PDLC}
\label{five}
When no control is applied, each air conditioner will operate according to its own duty cycle as described in Sec.\ref{three}. All the room temperatures are controlled around their respective set points, and the average room temperature is approximately equal to the average room set point, namely $T_{k}^{ave}\approx T_{set}^{ave}$. From the first theorem, the system will evolve into SSTE within a few steps when the PDLC is applied. We say that the system is in \textit{steady state} at time $k$ if it is in SSTE and $T_{k}^{i}\in(T_{min}^{i},T_{max}^{i})$ for all $i$. When the PDLC solution is applied in steady state, consumers in each room have the freedom to choose the set point to be whatever they want. After the set point is given, the operator will choose the comfort band for consumers around their preferred setting. The comfort band may be large or small, depending on the outside temperature and the energy we have purchased a day ahead. It is a compromise in the PDLC that consumers allow the operator to calculate the comfort band in order to achieve a smoother consumption. Denote the comfort band for room $i$ around $T_{set}^{i}$ by $(T_{min}^{i},T_{max}^{i})=(T_{set}^{i}-\Delta_{2},T_{set}^{i}+\Delta_{1})$ ($\Delta=\Delta_{1}+\Delta_{2}$ being a fixed value, namely we provide a fixed-valued comfort band for all the consumers). Define
\begin{equation}
\label{CRI}
T_{cr}^{i}=\frac{T_{max}^{i}-aT_{out}}{1-a},
\end{equation}
as the critical temperature point of room $i$. The physical meaning of $T_{cr}^{i}$ is the following: if room $i$'s temperature exceeds $T_{cr}^{i}$ at time $k$, then it needs packet at time $k$. Otherwise its room temperature will exceed $T_{max}^{i}$ at time $k+1$. The following two lemmas provide restrictions on how we choose $\Delta_{1}$ and $\Delta_{2}$. The first lemma provides a condition that the temperature of room $i$ will not exceed $T_{max}^{i}$ for all $i$, and the second lemma provides a condition that the temperature of room $i$ will not go below $T_{min}^{i}$ for all $i$.

\textbf{Lemma 1.} Assuming the system is in SSTE, and $T_{k^{\star}}^{i}\in(T_{min}^{i},T_{max}^{i})$ for all $i$ at time $k^{\star}$, if we provide $m$ packets, and $\Delta$ and $\Delta_{2}$ have been chosen to satisfy 
\begin{equation}
\label{LOW1}
\frac{\Delta_{2}}{\Delta}<\frac{m+1}{N_{c}},
\end{equation}
then there exists $\delta>0$ such that $T_{k^{\star}+1}^{i} < T_{max}^{i}$ for all $i$ with any packet length $\Delta t\in(0,\delta)$.

\textit{Proof:} If $T_{k^{\star}+1}^{r} \geq T_{max}^{i}$, then we have at least $m+1$ rooms with temperature beyond their critical point at time $k^{\star}$. Enumerate the $m+1$ (or more) consumers whose room temperature $T_{k^{\star}}^{i} \geq T_{cr}^{i}$ at time $k^{\star}$:$\textit{S}=\{i_{1},\cdots,i_{m+1}\}$. The remaining $N_{c}-m-1$ (or fewer) rooms' temperature are greater than $T_{min}^{i}$ for $i=m+2,\cdots,N_{c}$. The average room temperature \textit{lower bound} at time $k^{\star}$ is given by
\begin{equation}
\label{LOW2}
T_{k^{\star}}^{low}=\frac{1}{N_{c}}[\sum_{i_{j}\in\textit{S}}T_{cr}^{i_{j}}+\sum_{i_{j}\notin\textit{S}}T_{min}^{i_{j}}].
\end{equation}
We have
\begin{equation}
\label{LOW3}
\setlength\arraycolsep{0.1em}
\setlength{\extrarowheight}{7pt}
\begin{array}{cll}
T_{k^{\star}}^{low}-T_{k^{\star}}^{ave} & = & \frac{1}{N_{c}}[ \sum_{i_{j}\in\textit{S}}T_{cr}^{i_{j}}+\sum_{i_{j}\notin\textit{S}}T_{min}^{i_{j}}-\sum_{i=1}^{N_{c}}T_{set}^{i}] \\
 & =& \frac{1}{N_{c}}[\sum_{i_{j}\in\textit{S}}\frac{T_{max}^{i_{j}}-aT_{out}}{1-a}-\sum_{i_{j}\in\textit{S}}T_{set}^{i_{j}} \\ && -(N_{c}-m-1)\Delta_{2}] \\
& \propto  & [\sum_{i_{j}\in\textit{S}}T_{max}^{i_{j}}-(m+1)T_{out}]- \\
&&  [\sum_{i_{j}\in\textit{S}}T_{min}^{i_{j}}+N_{c}\Delta_{2}-(m+1)T_{out}]e^{-\frac{\Delta t}{\tau}}.
\end{array}
\end{equation}
The first equality is derived from $T_{k^{\star}}^{ave}=\frac{1}{N_{c}}\sum_{i=1}^{N_{c}}T_{set}^{i}$, namely at time $k^{\star}$ in SSTE the average room temperature is equal to the average temperature set point. The second equality is derived from $T_{min}^{i}=T_{set}^{i}-\Delta_{2}$ and (\ref{CRI}). The last proportionality is derived by plugging $a=1-e^{-\frac{\Delta t}{\tau}}$ from (\ref{DT}).\\
If we choose $\Delta$ and $\Delta_{2}$ to satisfy (\ref{LOW1}), then
\begin{equation}
\frac{(m+1)T_{out}-\sum_{i_{j}\in\textit{S}}T_{max}^{i_{j}}}{(m+1)T_{out}-\sum_{i_{j}\in\textit{S}}T_{min}^{i_{j}}-N_{c}\Delta_{2}}<1.
\end{equation}
Note that the above inequality is strict, so there exists $\delta>0$ such that
\begin{equation}
%\begin{array}{c}
\label{PACK}
\frac{(m+1)T_{out}-\sum_{i_{j}\in\textit{S}}T_{max}^{i_{j}}}{(m+1)T_{out}-\sum_{i_{j}\in\textit{S}}T_{min}^{i_{j}}-N_{c}\Delta_{2}}=e^{-\frac{\delta}{\tau}}.
%\end{array}
\end{equation}
Letting $\Delta t=\delta$ in (\ref{LOW3}), we have $T_{k^{\star}}^{low}=T_{k^{\star}}^{ave}$. Since (\ref{LOW3}) is monotonically decreasing as a function of $\Delta t$, then for packet length $\Delta t\in (0,\delta)$ we will have $T_{k^{\star}}^{low}-T_{k^{\star}}^{ave}>0$. Namely the average room temperature lower bound is greater than the average room temperature, which is a contradiction. We must have $T_{k^{\star}+1}^{i} < T_{max}^{i}$ for all $i$. $\blacksquare$

\textbf{Lemma 2.} Assuming the system is in SSTE, and $T_{k^{\star}}^{i}\in(T_{min}^{i},T_{max}^{i})$ for all $i$ at time $k^{\star}$, if we provide $m$ packets, and $\Delta$ and $\Delta_{1}$ have been chosen to satisfy 
\begin{equation}
\label{HIGH1}
\frac{\Delta_{1}}{\Delta}<\frac{N_{c}-m+1}{N_{c}},
\end{equation}
then there exists $\gamma>0$ such that $T_{k^{\star}+1}^{i} > T_{min}^{i}$ for all $i$ with packet length $\Delta t\in(0,\gamma)$.

\textit{Proof:} The proof is similar to lemma 1. We first assume that $T_{k^{\star}+1}^{r} \leq T_{min}^{i}$, then derive a average temperature upper bound $T_{k^{\star}}^{upp}$ at time $k^{\star}$ which is smaller than $T_{k^{\star}}^{ave}$ to show contradiction. We omit the details. $\blacksquare$

Based on the above two lemmas, we provide the following theorem for the steady state operation of the PDLC.

\textbf{Theorem 2.} Assuming that the system is in SSTE at time $k^{\star}$, and $T_{k^{\star}}^{i}\in(T_{min}^{i},T_{max}^{i})$ for all $i$, if we provide $m=s_{on}N_{c}$ number of packets over time and choose $\Delta_{1},\Delta_{2}$ such that
\begin{equation}
\label{THEO1}
\frac{\Delta_{1}}{\Delta}=\frac{N_{c}-m}{N_{c}}=s_{off},\frac{\Delta_{2}}{\Delta}=\frac{m}{N_{c}}=s_{on},
\end{equation}
then $T_{k}^{i}\in(T_{min}^{i},T_{max}^{i})$ for all $i$ and $k\geq k^{\star}+1$ with packet length $\Delta t\in(0,\min\{\delta,\gamma\})$.

\textit{Proof:} Clearly (\ref{THEO1}) satisfies (\ref{LOW1}) and (\ref{HIGH1}), and with packet length $\Delta t\in(0,\min\{\delta,\gamma\})$ both lemma 1 and lemma 2 will stand. We will have $T_{k^{\star}+1}^{i}\in (T_{min}^{i},T_{max}^{i})$ for all $i$. Since we provide $m=s_{on}N_{c}$ packets at time $k^{\star}$, the system is also in SSTE at time $k^{\star}+1$. By mathematical induction we can prove that $T_{k}^{i}\in(T_{min}^{i},T_{max}^{i})$ for all $i$ and $k\geq k^{\star}+1$. $\blacksquare$

\textbf{Remark 1.} As the comfort band $\Delta \rightarrow 0$, we have $\Delta_{1} \rightarrow 0, \Delta_{2} \rightarrow 0, T_{min}^{i}\approx T_{max}^{i},\forall i$. According to (\ref{PACK}) we must have $\Delta t \rightarrow 0$, which means we switch packets at increasingly large frequencies. In this case, individual room temperatures will stay at individual room set points after time $k>k^{\star}$ once $T_{k^{\star}}^{i}\approx T_{set}^{i}$ at time $k^{\star}$ for all $i$. This means that the width of the temperature band can be made to approach zero by letting the packet length approach zero. In actual implementation, there are practical limits on the minimum acceptable value of $\Delta t$, say 30 seconds or 1 minute, since the air conditioning unit cannot be switched on and off at an arbitrary frequency. Hence, convergence is to the comfort band and not to the actual set point.

\textbf{Remark 2.} From (\ref{THEO1}), $\Delta_{1}=s_{off}\Delta,\Delta_{2}=s_{on}\Delta$. When $s_{on}>s_{off}$, we have $\Delta_{2}>\Delta_{1}$. This can be explained by the intuition that since we are providing packets to more than a half number of consumers ($s_{on}>0.5$), it is more likely to have consumers being over-cooled. Thus we set a larger value of $\Delta_{2}$ to avoid such an occurrence. Similarly when $s_{on}<s_{off}$, we set a larger value of $\Delta_{1}$ to avoid consumers being over-warmed.

\textbf{Remark 3.} Based on the weather prediction, the building would purchase certain amount of packets a day ahead. In real time, the number of packets may not be enough if the predicted temperature is lower than what is actually realized. With the PDLC solution, the operator does not need to purchase additional energy from the real time market when the price is high. The operator can make packets switch more frequently to guarantee temperature control. In such cases, the average room temperature will converge to another value within the comfort band.

\textbf{Remark 4.} The packet length above is a theoretical value to guarantee temperature control in steady state. In the proof we focus on the worst case when initially at time $k^{\star}$ the temperatures of many rooms are in the vicinity of their maximum or minimum comfort boundary. In practice, the initial temperatures will be distributed more evenly across the comfort band. In such cases, the practical packet length can be larger than the theoretical value.

\begin{figure}[htb]
\centering
\includegraphics[width=0.4\textwidth,height=0.2\textheight]{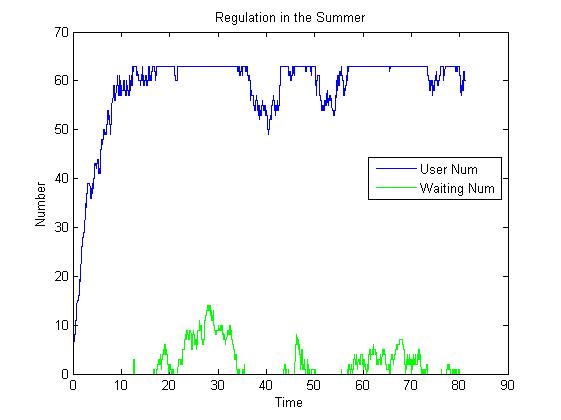}
\caption{Number of packets consumption and waiting consumers}
\label{binary}
\end{figure}

\textbf{Remark 5.} In our model we assume that the operator achieves all the temperature information within the building, and such information is continuous. In a companion technical report \cite{ZBW}, we assume that the operator must act on more restricted information. In this model, the appliance pool operator does not have complete and continuous access to appliance information, but instead receives requests for electricity that appliances send based on their own sensor readings. The operator receives packet request (withdrawal) from room $i$ when its room temperature reaches $T_{max}^{i}$ ($T_{min}^{i}$). The total number of available packets is limited which is equal to the expected average consumption. Packet supply is modelled as a multi-server queuing system with fixed service time (packet length). In a stochastic simulation, at certain times consumers have to wait to be served, and at other times the total number of packets cannot be fully used, see Fig.\ref{binary}. This indicates that continuous temperature information and control by an appliance pool operator results in a better control solution than binary information.

\section{From SSTE to Steady State}
\label{link section}
The final question is how we start from SSTE and find a packet allocation mechanism such that at time $k^{\star}$ we can start at $T_{k^{\star}}^{i}\in (T_{min}^{i},T_{max}^{i})$ for all $i$. According to the discrete time thermal dynamics,
\begin{equation}
\label{relation_k_kp1}
\setlength\arraycolsep{0.1em}
\setlength{\extrarowheight}{7pt}
\begin{array}{cll}
T_{k+1} & = & T_{k}+a(T_{out}-T_{k}-u_{k}T_{g}) \\
& = & T_{k}+(1-e^{-\frac{\Delta t}{\tau}})(T_{out}-T_{k}-u_{k}T_{g}) \\
& = & T_{k}+(1-(1-e^{-\frac{\Delta t}{\tau}}+o(\Delta t)))(T_{out}-T_{k}-u_{k}T_{g}) \\
& \approx & T_{k}(1-\frac{\Delta t}{\tau})+\frac{\Delta t}{\tau}(T_{out}-u_{k}T_{g}),
\end{array}
\end{equation}
where the third equality and fourth approximation are by Taylor series expansion for small packet length $\Delta t$. By a similar derivation we have,
\begin{equation}
\setlength\arraycolsep{0.1em}
\setlength{\extrarowheight}{7pt}
\begin{array}{cll}
T_{k+2} & \approx & T_{k+1}(1-\frac{\Delta t}{\tau})+\frac{\Delta t}{\tau}(T_{out}-u_{k}T_{g})\\
& = & T_{k}(1-\frac{2\Delta t}{\tau})+\frac{\Delta t}{\tau}(2T_{out}-(u_{k}+u_{k+1})T_{g}),
\end{array}
\end{equation}
where the second equality is obtained by plugging into (\ref{relation_k_kp1}) and ignoring terms of $o(\Delta t)$ for small $\Delta t$. For $N$ intervals, we have,
\begin{equation}
T_{k+N}= T_{k}(1-\frac{N\Delta t}{\tau})+\frac{\Delta t}{\tau}(NT_{out}-\sum\limits_{i=0}^{N}u_{k+i} T_{g}).
\end{equation}
Denote
\begin{equation}
n=\sum\limits_{i=0}^{N}u_{k+i}
\end{equation}
as the number of packets received within $N$ periods, then the temperature at time $t+N$ is given by,
\begin{equation}
\label{N_period_n_use}
T_{k+N}=T_{k}(1-\frac{N\Delta t}{\tau})+\frac{\Delta t}{\tau}(NT_{out}-nT_{g}).
\end{equation}
Having discussed the discrete time temperature evolution, we propose the following theorem to guarantee that if we start from SSTE, then there exists a packet allocation solution to satisfy the assumptions in theorem 2.

\textbf{Theorem 3.} If the aggregate system is in SSTE at time $k$ (per the conclusion of Theorem 1), let $n_{i}$ denote the number of packets received by room $i$ over the next $N$ successive time intervals of length $\Delta t$.  There exists a choice of packet allocation $\{n_{1},n_{2},\dots,n_{N_{c}}\}$ such that each room temperature is within the consumer's designated comfort band at time $k+N$.  That is, $T_{k+N}^{i}\in (T_{min}^{i},T_{max}^{i})$, with the total of allocated packets satisfying
\begin{equation}
\sum\limits_{i=1}^{N_{c}}n_{i}=mN.
\end{equation}

%If the system is in SSTE at time $k$, denote $n_{i}$ as the number of packet received by room $i$ in total $N$ discrete periods, then there exists a set of choice $\{n_{1},n_{2},\ldots,n_{N_{c}}\}$ such that all the room temperatures at time $k+N$ are within their respected comfort band, i.e. $T_{k+N}\in [T_{min}^{i},T_{max}^{i}]$, and that 
%\begin{equation}
%\sum\limits_{i=1}^{N_{c}}n_{i}=mN,
%\end{equation}
%namely the total packet consumption in $N$ periods are $mN$.
\textit{Proof.} According to (\ref{N_period_n_use}), after a total number of $n_{i}$ packet consumption in a successive $N$ periods starting at time $k$, the temperature in room $i$ at time $k+N$ is given by,
\begin{equation}
T_{k+N}^{i}=T_{k}^{i}(1-\frac{N\Delta t}{\tau})+\frac{\Delta t}{\tau}(NT_{out}-n_{i}T_{g}).
\end{equation}
The allowable choice of $n_{i}$ such that $T_{k+N}^{i}\in (T_{set}^{i}-\Delta_{1},T_{set}^{i}+\Delta_{2})$ is given by,
\begin{equation}
\label{ineq_ni}
\setlength\arraycolsep{0.1em}
\setlength{\extrarowheight}{7pt}
\begin{array}{l}
\frac{(T_{k}^{i}-T_{set}^{i}+\Delta_{1})\tau+N\Delta t(T_{out}-T_{k}^{i})}{\Delta t T_{g}}> n_{i}> \frac{(T_{k}^{i}-T_{set}^{i}-\Delta_{2})\tau+N\Delta t(T_{out}-T_{k}^{i})}{\Delta t T_{g}}.
\end{array}
\end{equation}
In order to have at least one integer $n_{i}$ within the bounds above, we need to have,
\begin{equation}
\setlength\arraycolsep{0.1em}
\setlength{\extrarowheight}{7pt}
\begin{array}{l}
\frac{(T_{k}^{i}-T_{set}^{i}+\Delta_{1})\tau+N\Delta t(T_{out}-T_{k}^{i})}{\Delta t T_{g}}-\frac{(T_{k}^{i}-T_{set}^{i}-\Delta_{2})\tau+N\Delta t(T_{out}-T_{k}^{i})}{\Delta t T_{g}}\geq 1 ,
\end{array}
\end{equation}
which can be achieved with a packet length
\begin{equation}
\Delta t \leq \frac{(\Delta_{1}+\Delta_{2}) \tau}{T_{g}}.
\end{equation}
We introduce the floor and ceil operator $\lfloor\cdot\rfloor$, $\lceil\cdot\rceil$. Let
\begin{equation}
\setlength\arraycolsep{0.1em}
\setlength{\extrarowheight}{7pt}
\begin{array}{l}
\alpha_{k}^{i}=\lfloor\frac{(T_{k}^{i}-T_{set}^{i}+\Delta_{1})\tau+N\Delta t(T_{out}-T_{k}^{i})}{\Delta t T_{g}}\rfloor,\\
\beta_{k}^{i}=\lceil\frac{(T_{k}^{i}-T_{set}^{i}-\Delta_{2})\tau+N\Delta t(T_{out}-T_{k}^{i})}{\Delta t T_{g}}\rceil,
\end{array}
\end{equation}
then $n_{i}$ can be chosen from integers between $\alpha_{k}^{i}$ and $\beta_{k}^{i}$. If the following inequality holds,
\begin{equation}
\label{alpha beta requirements}
\sum\limits_{i=1}^{N_{c}}\alpha_{k}^{i}\geq mN\geq \sum\limits_{i=1}^{N_{c}}\beta_{k}^{i},
\end{equation}
then there exists a choice of packet allocation $\{n_{1},n_{2},\dots,n_{N_{c}}\}$ such that (\ref{ineq_ni}) holds and
\begin{equation}
\sum\limits_{i=1}^{N_{c}}n_{i}=mN.
\end{equation}
Note that 
\begin{equation}
\begin{array}{cll}
\sum\limits_{i=1}^{N_{c}}\alpha_{k}^{i} & \geq & \sum\limits_{i=1}^{N_{c}}(\frac{(T_{k}^{i}-T_{set}^{i}+\Delta_{1})\tau+N\Delta t(T_{out}-T_{k}^{i})}{\Delta t T_{g}}-1)\\
& = & \frac{(\sum\limits_{i=1}^{N_{c}}T_{k}^{i}-\sum\limits_{i=1}^{N_{c}}T_{set}^{i}+N_{c}\Delta_{1})\tau+N\Delta t(N_{c}T_{out}-\sum\limits_{i=1}^{N_{c}}T_{k}^{i})}{\Delta t T_{g}} -Nc \\
& = & \frac{NN_{c}(T_{out}-T_{set}^{ave})}{T_{g}} + N_{c}(\frac{\Delta_{1}\tau}{\Delta t T_{g}}-1)\\
& = & mN + N_{c}(\frac{\Delta_{1}\tau}{\Delta t T_{g}}-1)\\
& \geq & mN, 
\end{array}
\end{equation}
and this holds as long as we choose $\Delta t$ such that $\frac{\Delta_{1}\tau}{\Delta t T_{g}}\geq 1$. In the derivation above, the third equality is obtained by the SSTE at time $k$ satisfying 
\begin{equation}
\sum\limits_{i=1}^{N_{c}}T_{k}^{i}=\sum\limits_{i=1}^{N_{c}}T_{set}^{i}=N_{c}T_{set}^{ave}.
\end{equation}
With similar derivation, a packet length $\Delta t$ such that $\frac{\Delta_{2}\tau}{\Delta t T_{g}}\geq 1$ will guarantee the second inequality in (\ref{alpha beta requirements}). To summarize, a packet length satisfying
\begin{equation}
\label{dt limit}
\Delta t \leq \min\{\Delta_{1},\Delta_{2}\}\frac{\tau}{T_{g}}
\end{equation}
will make (\ref{alpha beta requirements}) hold.
%If we choose $n_{i}$ such that
%\begin{equation}
%n_{i}=\frac{(T_{k}^{i}-T_{set}^{i})\tau+N\Delta t(T_{out}-T_{k}^{i})}{\Delta t T_{g}}, \forall i.
%\end{equation}
%Apparently it satisfies inequality (\ref{ineq_ni}) which guarantees that at time $k+N$ all the room temperatures are within their respected comfort band. Also we have,
%\begin{equation}
%\setlength\arraycolsep{0.1em}
%\setlength{\extrarowheight}{7pt}
%\begin{array}{cll}
%\sum\limits_{i=1}^{N_{c}}n_{i} & = & \sum\limits_{i=1}^{N_{c}}\frac{(T_{k}^{i}-T_{set}^{i})\tau+N\Delta t(T_{out}-T_{k}^{i})}{\Delta t T_{g}}\\
%&=& \frac{(\sum\limits_{i=1}^{N_{c}}T_{k}^{i}-\sum\limits_{i=1}^{N_{c}}T_{set}^{i})\tau+N\Delta t(N_{c}T_{out}-\sum\limits_{i=1}^{N_{c}}T_{k}^{i})}{\Delta t T_{g}}\\
%&=&\frac{NN_{c}(T_{out}-T_{set}^{ave})}{T_{g}}\\
%&=&mN
%\end{array}
%\end{equation}
%where the third equality is by the SSTE at time $k$ satisfying 
%\begin{equation}
%\sum\limits_{i=1}^{N_{c}}T_{k}^{i}=\sum\limits_{i=1}^{N_{c}}T_{set}^{i}=N_{c}T_{set}^{ave}
%\end{equation}
This ends the proof of theorem 3. $\blacksquare$

\textbf{Remark.} According to (\ref{dt limit}), the upper bound of packet length is directly proportional to $\tau$ and inversely proportional to $T_{g}$. The intuition is that large value of $\tau$ impedes and $T_{g}$ facilitates the thermal transmission, which allows larger and requires smaller packet length respectively.

The remaining issue is to assign $m$ packets at each period. Denote $a_{i,k}$ as the binary variable representing packet assignment at time $k$ for room $i$. Up to time $k+j$, define
\begin{equation}
\label{dyn_need_packets}
n_{i}(k+j)=n_{i}-\sum\limits_{l=k}^{k+j}a_{i,l}
\end{equation}
as the remaining number of packet needed for room $i$ until time $k+N$. A simple allocation algorithm works as follows, starting at time $k$ we allocate packets to the $m$ rooms with largest $n_{i}(k)$. Let $a_{i,k}=1$ if packet is allocated and $0$ otherwise. Use (\ref{dyn_need_packets}) to update $n_{i}(k+1)$ for all $i$. Repeating such allocation procedure until the end of interval $k+N$ will guarantee $m$ allocation each period. 

We first prove the following inequality of $n_{i}(k+j)$, 
\begin{equation}
\label{rangeofni}
0 \leq n_{i}(k+j) \leq N-j.
\end{equation}
We prove with induction. Note that for $j=l=0$ is it apparently true. Also at time $k+l$,
\begin{equation}
\label{mN1}
\sum\limits_{i=1}^{N_{c}}n_{i}(k+l)=mN-\sum\limits_{j=0}^{l}\sum\limits_{i=1}^{N_{c}}a_{i,k+j}=m(N-l).
\end{equation}
For $j=l+1$, we proof with contradiction. If there exists a room $i^{\star}$ such that $n_{i^{\star}}(k+l)\leq N-l$ and $n_{i^{\star}}(k+l+1) > N-l-1$, then $n_{i^{\star}}(k+l) = N-l$. It also indicates that room $i^{\star}$ does not get a packet and there are at least $m$ rooms, indexed by $i^{j},j=1,\ldots,m$, other than $i^{\star}$ such that $n_{i^{j}}(k+l) = N-l$ to get packets. Then 
\begin{equation}
\setlength\arraycolsep{0.1em}
\setlength{\extrarowheight}{7pt}
\begin{array}{cll}
\sum\limits_{i=1}^{N_{c}}n_{i}(k+l) & \geq & \sum\limits_{j=1}^{m}n_{i^{j}}(k+l)+n_{i^{\star}}(k+l)\\
& = & (m+1)(N-l),
\end{array}
\end{equation}
which contradicts (\ref{mN1}). So we will have $n_{i}(k+l+1) \leq N-l-1$ for $j=l+1$ and all $i$. 

To show that $n_{i}(k+l+1) \geq 0$ for all $i$. Suppose that $n_{i^{\star}}(k+l+1) < 0$, it indicates that $n_{i^{\star}}(k+l) = 0$ and room $i^{\star}$ gets a packet. Thus there are at most $m-1$ rooms, indexed by $i^{j},j=1,\ldots,m-1$, with positive value of $n_{i^{j}}(k+l)>0$. Then
\begin{equation}
\setlength\arraycolsep{0.1em}
\setlength{\extrarowheight}{7pt}
\begin{array}{cll}
\sum\limits_{i=1}^{N_{c}}n_{i}(k+l) & = & \sum\limits_{j=1}^{m-1}n_{i^{j}}(k+l)+n_{i^{\star}}(k+l)\\
& \leq & (m-1)(N-l),
\end{array}
\end{equation}
contradicting (\ref{mN1}) again. So we will have $n_{i}(k+l+1) \geq 0$ for $j=l+1$ and all $i$. Using mathematical induction, for all $i=1,\ldots,N_{c}$ and $j=0,\ldots,N$, (\ref{rangeofni}) holds. Then for $j=N$ and all $i$, we will have 
\begin{equation}
n_{i}(k+N)=0.
\end{equation}
Namely all the rooms will have received the exact packets they need and $T_{k+N}^{i}\in (T_{min}^{i},T_{max}^{i})$ for all $i$ with $m$ packets allocation for each period. The intuition of such allocation is to provide packets to the $m$ rooms that have largest temperature deviation above their target, namely at time $k+j$ the $m$ rooms with largest $n_{i}(k+j)$ receive packet for $j=0,\ldots,N$. 

To summarize, theorem 1 guarantees that the systems will evolve into SSTE, theorem 3 guarantees that starting from SSTE we have an allocation solution such that we can have $T_{k^{\star}}^{i}$ within the comfort band of room $i$ for all $i$, and theorem 2 guarantees temperature control after the allocation. The three theorems complete the overall PDLC mechanism.

\section{Robustness Analysis of the PDLC}
\label{six}
While Ihara and Schweppe's model \cite{IS} is deterministic, we have also considered temperature disturbances to get a thermal model that reflects uncertainty. Temperature disturbance in real life may come with the inaccuracy of sensors, the unpredictability of consumers, etc. The revised temperature dynamics is therefore given by
\begin{equation}
\label{RCT}
\frac{dT}{dt}=\frac{T_{out}-T-T_{g}u+\epsilon(t)}{\tau},
\end{equation}
where $\epsilon(t)$ is a bounded thermal stochastic disturbance uniformly distributed between $[-\bar{\epsilon},\bar{\epsilon}]$. We investigate the transient and steady state operation of the PDLC solution under this model of disturbance to illustrate the robustness of the PDLC. The discrete version of the model becomes
\begin{equation}
T_{k+1}=(1-a)T_{k}+aT_{out}-aT_{g}u_{k}+a\epsilon_{k}.
\end{equation}
Repeating the derivation in theorem 1, the average room temperature evolution from time $k$ to $k+1$ given by
\begin{equation}
T_{k+1}^{ave}=T_{k}^{ave}+a(T_{set}^{ave}-T_{k}^{ave})+\frac{a}{N_{c}}\sum_{i=1}^{N_{c}}\epsilon_{k}^{i}.
\end{equation}
Note that the term $\frac{a}{N_{c}}\sum_{i=1}^{N_{c}}\epsilon_{k}^{i}$ is bounded between $[-a\bar{\epsilon},a\bar{\epsilon}]$. When packet length $\Delta t$ is small, $a$ will approach zero, and this makes the disturbance term approach zero. Then the average room temperature will still converge to $T_{set}^{ave}$.

As for the steady state operation, we will have the same comfort band selection as in Theorem 2, namely $\Delta_{1}=s_{off}\Delta,\Delta_{2}=s_{on}\Delta$ with the difference in the boundary of packet length selection. In the model with disturbances, we can similarly derive the contingent packet length $\delta^{'}$ and $\gamma^{'}$ as in lemma 1 and 2. For example, the value of $\delta^{'}$ will satisfy
\begin{equation}
\begin{array}{c}
\frac{(m+1)(T_{out}+\bar{\epsilon})-\sum_{i_{j}\in\textit{S}}T_{max}^{i_{j}}}{(m+1)(T_{out}+\bar{\epsilon})-\sum_{i_{j}\in\textit{S}}T_{min}^{i_{j}}-N_{c}\Delta_{2}}=e^{-\frac{\delta^{'}}{\tau}}.
\end{array}
\end{equation}
Compared with (\ref{PACK}), the only difference is that the term $T_{out}$ in (\ref{PACK}) is replaced by $T_{out}+\bar{\epsilon}$. Hence the disturbance in (\ref{RCT}) can be understood as the uncertainty introduced by the outside temperature. Also, the above $\delta^{'}$ is smaller than the $\delta$ in lemma 1. This is no surprise since the existence of uncertainty forces us to switch packets more frequently.

\section{Simulation}
\label{seven}
\subsection{Air Conditioner Temperature Control}
\begin{figure}[hbt]
\centering
\subfigure[Warm load pickup process with the PDLC]{
\includegraphics[width=0.4\textwidth,height=0.2\textheight]{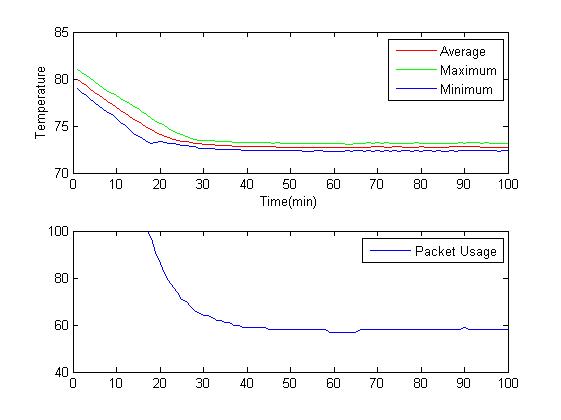}
\label{WP2}
}
\subfigure[Warm load pickup process without control]{
\includegraphics[width=0.4\textwidth,height=0.2\textheight]{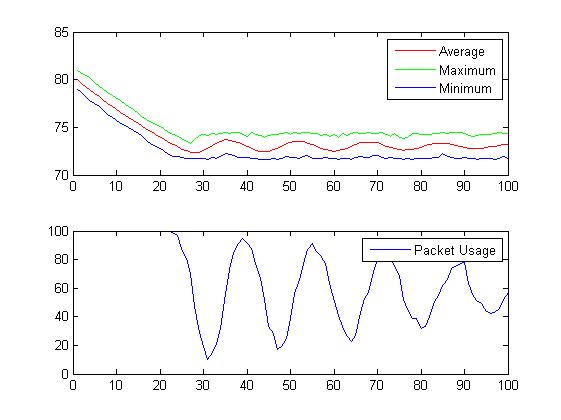}
\label{WP3}
}
\label{WP}
\caption{Loads start outside the comfort zone}
\end{figure}

We simulate air conditioner temperature control process to verify theoretical results. Environmental parameters are $T_{g}=40, T_{out}=93, \tau=20, N_{c}=100, \bar{\epsilon}=10$. Consumers preferred set point is  $T_{set}^{i}=73$ for all $i$. After calculation we choose $T_{max}=74,T_{min}=72,\Delta t=1$. Fig.3 is the process of warm load pick up. Fig.\ref{WP2} shows that the average room temperature converges to the set point when we applied the number of packets at time $k$ as a function of $T_{k}^{ave}-T_{set}^{ave}$, which verifies theorem 1. Compared with Fig.\ref{WP3} where no control is applied, the consumption oscillation by the PDLC solution is reduced by a large amount after the system evolves into SSTE. The oscillation magnitude in Fig.\ref{WP3} continues to exist if we simulate for longer time.
Fig.4 is the steady state process where all the rooms have their initial temperatures randomly distributed within their comfort bands. We see two main advantages of our PDLC solution. First, the maximum and minimum room temperature are controlled within the comfort band in steady state, which cannot be achieved without control since then the disturbance drives the temperature outside the comfort band. Second, the consumption process is smoother with PDLC solution than in the stochastic uncontrolled case.

\begin{figure}[hbt]
\centering
\subfigure[Steady state operation with the PDLC]{
\includegraphics[width=0.4\textwidth,height=0.2\textheight]{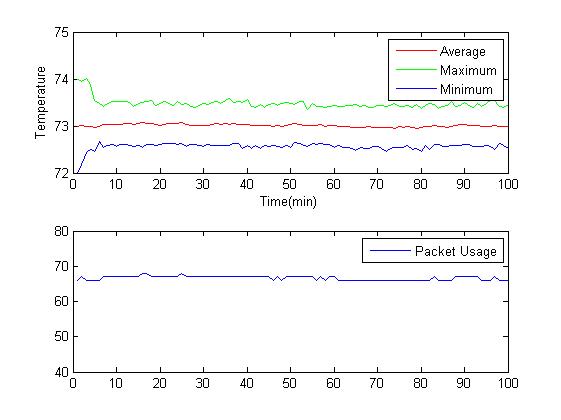}
\label{SS1}
}
\subfigure[Steady state operation without control]{
\includegraphics[width=0.4\textwidth,height=0.2\textheight]{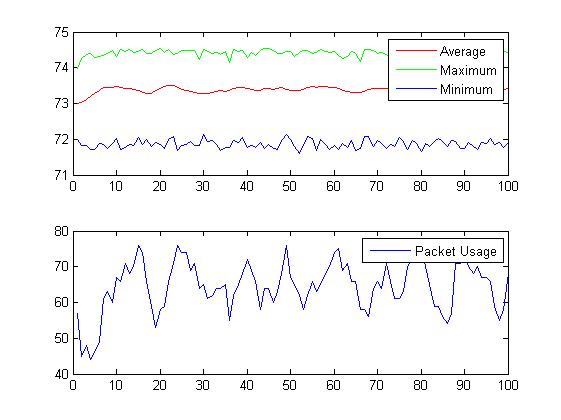}
\label{SS2}
}
\label{SS}
\caption{Loads start in the comfort zone}
\end{figure}
\subsection{Multiple Appliances Simulation}
Consider the simulation of multiple appliances. The controllable thermostatic loads are air conditioners and refrigerators. We also add uncontrollable loads, such as lighting and plug-in devices. The thermal characteristics of the refrigerator is similar to the air conditioner. Refrigerator parameters is given by $T_{set}=35,T_{g}=75,T_{out}=73,\tau=185$, we choose $T_{max}=38,T_{min}=32$. $t_{on}$ and $t_{off}$ are around 20 minutes according to (\ref{TOFF}) and (\ref{TON}), which is typical duty cycle of refrigerator \cite{CM}. We assume there are 60 refrigerators each consuming around 600 watts of power. The air conditioner consumes around $3kW$ each. There is also an industrial chiller that consumes with small variation in steady state, which is uniformly distributed between $[135,145]kW$. Uncontrollable loads are uniformly distributed between $[180,200]kW$. Table.1 shows the comparison result between the PDLC solution and the case when no control is applied. We find that standard deviation of consumption by the PDLC solution is nearly half of that without control. Also the maximum electric usage is reduced nearly $50\%$ from above its average.
\begin{table}
\caption{Comparison of Consumption Statistics}
\centering
\begin{tabular}{|c|c|c|c|c|}
\hline  & Mean & Std Dev & Maximum & Minimum \\ 
\hline PDLC & 725.86 & 8.18 & 744.09 & 709.92 \\ 
\hline No Control & 724.11 & 15.06 & 761.43 & 687.23 \\
\hline 
\end{tabular} 
\end{table}

\section{Conclusions and Future Work}
\label{eight}
This paper proposes an innovative PDLC solution for demand side management. We have discussed a thermal dynamic model of typical thermostatic appliances and derived a mathematical expression of its duty cycle. Three theorems are proposed to illustrate overall PDLC solution. The first theorem proves the convergence of the average room temperature to average room set point. The second theorem provides comfort band choice such that we can guarantee effective temperature control in steady state. The third theorem builds the bridge between the first two theorems. Simulation shows that the PDLC solution can provide comfortable temperature control with minimum consumption oscillation, and reduce consumption peaks at the same time. 

Future research will compare the performance of the PDLC as described here with comparable distribution control approaches using market based signaling. Renewable energy sources will be included, and the dynamics of an appliance pool operator buying and selling resources under different communication protocols will be studied.

\addtolength{\textheight}{-12cm}   % This command serves to balance the column lengths

%%%%%%%%%%%%%%%%%%%%%%%%%%%%%%%%%%%%%%%%%%%%%%%%%%%%%%%%%%%%%%%%%%%%%%%%%%%%%%%%

%%%%%%%%%%%%%%%%%%%%%%%%%%%%%%%%%%%%%%%%%%%%%%%%%%%%%%%%%%%%%%%%%%%%%%%%%%%%%%%%

%%%%%%%%%%%%%%%%%%%%%%%%%%%%%%%%%%%%%%%%%%%%%%%%%%%%%%%%%%%%%%%%%%%%%%%%%%%%%%%%
%\section*{APPENDIX}


%\section*{ACKNOWLEDGMENT}

%The preferred spelling of the word �acknowledgment� in America is without an �e� after the �g�. Avoid the stilted expression, �One of us (R. B. G.) thanks . . .�  Instead, try �R. B. G. thanks�. Put sponsor acknowledgments in the unnumbered footnote on the first page.

%%%%%%%%%%%%%%%%%%%%%%%%%%%%%%%%%%%%%%%%%%%%%%%%%%%%%%%%%%%%%%%%%%%%%%%%%%%%%%%%

%eferences are important to the reader; therefore, each citation must be complete and correct. If at all possible, references should be commonly available publications.

\end{document}